\documentclass[twocolumn,notitlepage,nofootinbib,superscriptaddress,longbibliography]{revtex4-1} 
\usepackage{latexsym}
\usepackage{graphicx}
\usepackage{amsmath}
\usepackage{mathrsfs}
\usepackage[normalem]{ulem} 
\usepackage{color}

\begin{document}

\title{Breakdown of Landau Fermi liquid theory: restrictions on the degrees of freedom of quantum electrons}

\author{Yuehua Su}
\email{suyh@ytu.edu.cn}
\affiliation{ Department of Physics, Yantai University, Yantai 264005, P. R. China}

\author{Hantao Lu}
\email{luht@lzu.edu.cn}
\affiliation{ Center for Interdisciplinary Studies, Lanzhou University, Lanzhou 730000, P. R. China}

\begin{abstract}

One challenge in contemporary condensed matter physics is to understand unconventional electronic physics beyond the paradigm of Landau Fermi-liquid theory. Here, we present a perspective that posits that most such examples of unconventional electronic physics stem from restrictions on the degrees of freedom of quantum electrons in Landau Fermi liquids. Since the degrees of freedom are deeply connected to the system's symmetries and topology, these restrictions can thus be realized by external constraints or by interaction-driven processes via the following mechanisms: (i) symmetry breaking, (ii) new emergent symmetries, and (iii) nontrivial topology. Various examples of unconventional electronic physics beyond the reach of traditional Landau Fermi liquid theory are extensively investigated from this point of view. Our perspective yields basic pathways to study the breakdown of Landau Fermi liquids and also provides a guiding principle in the search for novel electronic systems and devices. 

\end{abstract}

\maketitle


\section{Introduction}\label{sec1}

In the 1950s, L. P. Landau established a phenomenological theory for the low-energy physics of Fermi liquids in metals, which has now become widely known as Landau Fermi liquid theory~\cite{Landau1956,Landau1957,Landau1958}. Several principles are assumed to hold in Landau Fermi liquid theory. The first assumption is the Pauli exclusion principle and its associated Fermi--Dirac statistics. This assumption places extreme restrictions on the phase space of the low-energy fermionic particles near the Fermi energy. A second assumption of the theory is the adiabatic and analytic principle, which states that when interactions are turned on adiabatically in a non-interacting Fermi gas, the Fermi gas evolves into a Fermi liquid adiabatically and analytically without any phase transition. This smooth evolution results in a one-to-one correspondence between the free Fermi particles in the Fermi gas and the {\em quasiparticles} in the Fermi liquid. Landau then assumed that the low-energy physics of the Fermi liquid was dominated by the quasiparticles, and that the free energy could be described by
\begin{equation}
F[n] = E[n] - T S[n] , \label{eqn1.1}
\end{equation} 
where the energy $E[n]$ and entropy $S[n]$ are given by
\begin{eqnarray}
E[n] &=& E_0 + \sum_{\mathbf{k}\sigma} \varepsilon_{\mathbf{k}\sigma}^{(0)} 
\delta n_{\mathbf{k}\sigma} + \frac{1}{2}\sum_{\mathbf{k}\sigma;\mathbf{k}^{\prime}\sigma^{\prime}} 
f_{\mathbf{k}\sigma, \mathbf{k}^{\prime}\sigma^{\prime}}
\delta n_{\mathbf{k}\sigma} \delta n_{\mathbf{k}^{\prime}\sigma^{\prime}} , \nonumber \\
S[n] &=& - k_B \sum_{\mathbf{k}\sigma}  \left( n_{\mathbf{k}\sigma}\ln n_{\mathbf{k}\sigma} + 
(1-n_{\mathbf{k}\sigma})\ln (1-n_{\mathbf{k}\sigma}) \right) . \nonumber 
\end{eqnarray}
Here, $\varepsilon_{\mathbf{k}\sigma}=\frac{\delta E[n]}{\delta n_{\mathbf{k}\sigma}}$ is the energy of a quasiparticle with momentum $\mathbf{k}$ and spin $\sigma$, and 
$n_{\mathbf{k}\sigma}=\frac{1}{e^{(\varepsilon_{\mathbf{k}\sigma}-\mu_F)/T}+1}$ is the Fermi--Dirac function. A particular point to note is that only the forward scattering interactions are assumed to be involved in Landau Fermi liquid theory. 

Landau's theory can account for many universal properties of low-energy Landau Fermi liquids~\cite{Colemanbook}, including the well-defined Fermi surface, the linear-$T$ specific heat, the 
Pauli spin susceptibility, and the zero sounds. The highly successful phenomenological theory has also been firmly reestablished using quantum field theory~\cite{Abrikosovbook} and renormalization group 
theory~\cite{Shankar1994}, with the latter method showing the stability of Landau Fermi liquids in two- and three-dimensional (2D and 3D) spatial spaces, where the Landau forward scattering interactions are marginal.  

In the last several decades, much attention in the field of condensed matter has turned to the study of unconventional electronic physics beyond the Landau Fermi liquid framework~\cite{VarmaPhysRep2002,StewartNFLRMP2001}. Typical examples include one-dimensional (1D) Luttinger liquids~\cite{Voit1994}, disordered electrons with Anderson localization~\cite{AbrahamsBook}, heavy fermion superconductors~\cite{StewartNFLRMP2001,ColemanReview2015}, 
high-$T_c$ cuprate superconductors~\cite{KeimerNature2015,NormanCuSC2014,AndersonRVB2004,PALeeRMP2006}, Fe-based superconductors~\cite{ChenDaiFeSCs2014,StewartFeSCRMP2011}, and 2D electrons with quantum Hall effects~\cite{klitzing80,tsui82}. 
Some of the unconventional electronic physics evident in these systems include: (i) universal 1D Luttinger liquids, (ii) electron Anderson localization and the metal--insulator transition, (iii) novel topological structure of the bulk and surface states, (iv) unconventional superconductivity and novel mother normal states, and (v) quantum phase transitions and quantum criticality. 

While the theory behind 1D Luttinger liquids~\cite{Haldane1981,Voit1994} and the quantum Hall effects~\cite{prange2012,jainbook} has been well established, and despite Landau's theory of spontaneous symmetry breaking~\cite{Landaubook} and Wilson's renormalization group theory~\cite{WilsonRG1974} providing a good account of classical continuous phase transitions and the corresponding critical phenomena, the description of many unconventional electronic physical phenomena remains beyond these well-established formulations. Such unexplained phenomena include unconventional superconductivity, various novel normal states, quantum phase transitions, and quantum criticality. 

In this article, we present a new perspective on unconventional electronic physics beyond the paradigm of traditional Landau Fermi liquid theory. Our perspective is that most such unconventional electronic physics stems from the restriction of the degrees of freedom of the quantum electrons in Landau Fermi liquids. The degrees of freedom describe the independent ``motions" of quantum electrons and can be defined by a complete set of commuting observables that are deeply related to the symmetries of the quantum electrons. The degrees of freedom may also involve the nontrivial topological structure of the quantum electronic states, and thus, the degrees of freedom are deeply connected to both the symmetry and the topology of the system. This leads us naturally to the following proposal: the restriction of the degrees of freedom can be used to explain novel electronic physics beyond the standard Landau Fermi liquid theory, and such restrictions can be realized by external constraints or by interaction-driven dynamical processes via the mechanisms of (i) symmetry breaking, (ii) the emergence of new symmetries, and (iii) nontrivial topology. This perspective enables us to study the breakdown of Landau Fermi liquid theory via the restricted degrees of freedom of the quantum electrons, and can also guide us in the search for novel electronic systems and devices through manual control of the electronic degrees of freedom.  

The article is arranged as follows. In Section \ref{sec2}, we show how to specify the quantum electronic states. This leads to the definition of the degrees of freedom of quantum electrons, which physically, are 
deeply connected to the symmetry and topology of the system. The principles governing the novel electronic physics are also presented in this section. In the following sections, extensive reviews of various examples of unconventional electronic physics beyond the Landau Fermi liquid theory framework are provided from our perspective. Section \ref{sec3} focuses on the restriction of the spatial degrees of freedom, which can result in critical phenomena, 1D universal Luttinger liquids, and impurity-induced Anderson localization. Section \ref{sec4} considers the restriction of the internal degrees of freedom, such as the spin, charge, 
phase, and local basis states, which can lead to quantum Hall effects, spin- and charge-density waves (SDWs and CDWs), Pomeranchuk nematicity, superconductivity, and Mott--Hubbard physics. Section \ref{sec5} focuses on restrictions imposed by external degrees of freedom, which include the 
Goldstone bosons arising from continuous symmetry breaking, background fluctuations in Anderson's hidden Fermi liquid, Bose unparticles in the strange metallic state, and local magnetic moments in Kondo physics and heavy fermion superconductors. Finally, in Section \ref{sec6}, we present a summary of the possible causes of the breakdown of the Landau Fermi liquid framework in novel electronic physics, and suggest some proposals for further research.

\section{Breakdown of Landau Fermi liquid theory: basic ideas} \label{sec2}

\subsection{Degrees of freedom and basic principles} \label{sec2.1}

A degree of freedom is an independent physical variable in the formal description of the state of a physical system, which describes its independent ``motion".\footnote{From Wikipedia on the ``degrees of freedom".}
For example, a classical electron with mass $m$ and charge $-e$ is described by its position $(\mathbf{r})$ and momentum $(\mathbf{p})$ variables, which are functions of the absolute time parameter $t$. Thus, $\left\{ \mathbf{r}(t), \mathbf{p}(t)\right\}$ are the degrees of freedom of a classical electron.  In contrast, in the quantum electron case, the basic variables are quantum states and these states are defined in Hilbert space. Any quantum state in the Hilbert space can be expressed as $\left\vert \Psi(t) \right\rangle = \sum_{\alpha} \psi_{\alpha} (t) \left\vert \alpha \right\rangle$, where $ \left\vert \alpha \right\rangle $ are the basis states of the Hilbert space, and $\psi_{\alpha}(t)\equiv   \left\langle \alpha \vert \Psi(t) \right\rangle$ are quantum wavefunctions which can be quantized to form quantum fields in second quantization formalisms. Following Dirac~\cite{DiracBook}, the basis states $ \left\vert \alpha \right\rangle $ can be defined as the eigenfunctions of a complete set of commuting observables with their eigenvalues $\alpha $ being the so-called quantum numbers. The degrees of freedom of a quantum electron (which describe its independent ``motions") can then be defined by the complete set of commuting observables. Here, the commuting observables whose eigenfunctions define the basis states play a similar role to the independent variables which were introduced to identify the classical physical states. Mathematically, the basis states $ \left\vert \alpha \right\rangle $ can be defined as the irreducible representations of the symmetry group associated with the quantum electron~\cite{Weinberg2013}, 
while the complete set of commuting observables can be constructed from the generators of the symmetry group, with the quantum numbers $\alpha$ being the indices of the irreducible representations. Thus, the degrees of 
freedom of quantum electrons are deeply connected to the overall symmetry, where in general, the symmetry group involves the symmetries of the spatial-temporal space $\left( \mathbf{r}, t\right)$, and of the internal spaces, such as the spin and phase spaces. 

The quantum states of many-body quantum electrons can be similarly defined as
$\left\vert \Psi^{(N)}(t) \right\rangle = \sum_{\alpha_i} \psi_{\alpha_1 \alpha_2 \cdots \alpha_N} (t) \left\vert \alpha_1 \alpha_2 \cdots \alpha_N \right\rangle$, where $ \left\vert \alpha_1 \alpha_2 \cdots \alpha_N \right\rangle $ are the $N$-body basis states, and $\psi_{\alpha_1 \alpha_2 \cdots \alpha_N}(t) \equiv \left\langle \alpha_1 \alpha_2 \cdots \alpha_N 
\vert \Psi^{(N)}(t) \right\rangle$ are the $N$-body quantum wavefunctions. Here, $ \left\vert \alpha_1 \alpha_2 \cdots \alpha_N \right\rangle =\hat{\psi}^{\dag}_{\alpha_N} \cdots \hat{\psi}^{\dag}_{\alpha_2} \hat{\psi}^{\dag}_{\alpha_1} \left\vert 0 \right\rangle $, where the creation 
operators $\hat{\psi}^{\dag}_{\alpha_i}$ are the quantized fermionic fields of the corresponding wavefunctions of the $i$-th quantum electron. 
The degrees of freedom of the many-body quantum electrons can then be defined by the collection of the complete set of commuting observables of all the quantum electrons, with the eigenfunctions denoted by the set of quantum numbers $\{ \alpha_1 \alpha_2 \cdots \alpha_N \}$. As before, they are deeply related to the governing symmetry group of the many-body quantum electrons.

In some cases, the quantum states may have a global nontrivial topological structure that is out of reach of the usual symmetry-based description. Examples include vortices, quantum Hall states~\cite{thouless82,avron2003}, 
and topological insulators~\cite{SchnyderPRB2008,kitaev2009,HasanKane2010,QiZhangRMP2011,Chiu2016}. In these cases, besides the degrees of freedom defined by the 
symmetry, to identify a quantum state we should also specify the nature of its global topological structure. Thus, the topology of the system provides quantum electrons with an additional degree of freedom. Mathematically, topology is a special global invariant property of the quantum states, where the global invariance is defined in the spatial-temporal space or in the internal space.

As mentioned in the Introduction, a Landau Fermi liquid is assumed to evolve adiabatically and analytically from a non-interacting Fermi gas~\cite{Landau1956,Colemanbook}. This adiabatic and analytic evolution leads to a one-to-one correspondence between the ground and excited states of the Landau Fermi liquid and that of the Fermi gas. It also implies a one-to-one correspondence between the electrons in the Landau Fermi liquid 
and the electrons of the Fermi gas, the former of which are known as quasiparticles and which have conserved quantum numbers of momentum $\mathbf{k}$, spin $S=\frac{1}{2}$, and charge $-e$. These quantum numbers 
describe the translation, spin, and charge degrees of freedom and stem from the spatial translational symmetry, the spin $SU(2)$ symmetry, and the charge $U(1)$ symmetry of the system, respectively. The momentum operator $\mathbf{P}$, the spin operator $\mathbf{S}$, and the particle-number operator $N=-i\partial_{\theta}$ are the corresponding observable operators used to define these degrees of freedom.
In principle, the spin degree of freedom of electrons and the spin $SU(2)$ symmetry are fundamentally associated with the Lorentz symmetry of spacetime~\cite{AZee2016}. 
Generally, Landau Fermi liquids in crystalline condensed matter have space-group symmetries which include both spatial translational symmetry and point-group symmetries, the latter of which involves rotational symmetry, inversion symmetry, etc. These point-group symmetries thus define rotation and 
parity degrees of freedom, along with the associated rotational $L$ and parity $\mathscr{P}$ observable operators. Whether the system exhibits a continuous or discrete rotational symmetry has an important consequence on the topology of the Fermi surface as the spontaneous breaking of rotational symmetry can lead to Pomeranchuk electronic nematicity~\cite{Pomeranchuk1959}. 
In most Landau Fermi liquids, there are also the time translation and time reversal degrees of freedom, which are related to symmetry under time-translation and time-reversal. In such cases, the Hamiltonian $H$ and 
the time-reversal operator $\mathscr{T}$ are the relevant observable operators. Finally, as identical quantum particles, Landau quasiparticles also have an exchange degree of freedom and follow Fermi--Dirac statistics with permutation symmetry. This exchange symmetry places severe constraints on Landau Fermi liquids, with one of the consequences being the formation of the Fermi surface. Essentially, only low-energy quasiparticles near the Fermi surface are involved in the physics of the Landau Fermi liquids. 

As discussed above, the Landau Fermi-liquid ground state evolves adiabatically and analytically from a non-interacting Fermi gas~\cite{Landau1956,Colemanbook}. This can be described mathematically as 
\begin{equation}
\left\vert \Psi^{(FL)}_G \right\rangle = \hat{U} \left\vert \Psi^{(FL)}_{G,0} \right\rangle ,
\label{eqn2.1.1}
\end{equation}
where
$
\left\vert \Psi^{(FL)}_{G,0}  \right\rangle  \equiv \prod_{\left\vert \mathbf{k} \right\vert <k_F} 
c^{\dag}_{\mathbf{k}\uparrow} c^{\dag}_{\mathbf{k}\downarrow}\left\vert 0  \right\rangle 
$
is the ground state of the Fermi gas, $\hat{U}$ is the time-evolution operator with the Landau interactions included, and the Landau Fermi liquid ground state preserves all the above-mentioned symmetries. There are then two main types of low-energy excitations we must consider. One type are the quasiparticles defined by~\cite{Colemanbook}  
\begin{equation}
\left\vert \Psi^{(FL)}_{N+1} \right\rangle = \hat{U} \left( c^{\dag}_{\mathbf{p},\sigma}
\left\vert \Psi^{(FL)}_{G,0} \right\rangle \right) , \label{eqn2.1.2}
\end{equation}
which can be re-expressed as 
$
\left\vert \Psi^{(FL)}_{N+1} \right\rangle = \left( a^{\dag}_{\mathbf{p},\sigma} 
\left\vert \Psi^{(FL)}_{G} \right\rangle \right)
$
using the quasiparticle operators 
$a^{\dag}_{\mathbf{p},\sigma}\equiv \hat{U} c^{\dag}_{\mathbf{p},\sigma} \hat{U}^{\dag}$.
The other type of excitation are collective excitations, such as the CDW and SDW excitations, which can be 
defined by
\begin{equation}
\left\vert \Psi^{(FL)}_{O} \right\rangle = \hat{U} \left( \hat{O}_{\mathbf{q}}
\left\vert \Psi^{(FL)}_{G,0} \right\rangle \right).   \label{eqn2.1.3}
\end{equation}
For CDWs, $\hat{O}_{\mathbf{q}}$ is defined as 
\begin{equation}
n_{\mathbf{q}} = \sum_{i} n_i e^{i\mathbf{k}\cdot \mathbf{r}_i} =\sum_{\mathbf{k}\sigma} 
c_{\mathbf{k+q}\sigma}^{\dag} c_{\mathbf{k}\sigma}, \label{eqn2.1.4}
\end{equation}
while for SDWs, it is defined by 
\begin{equation}
\mathbf{S}_{\mathbf{q}} = \sum_{i} \mathbf{S}_i e^{i\mathbf{q}\cdot \mathbf{r}_i} =\sum_{\mathbf{k}\sigma\sigma^{\prime}} 
c_{\mathbf{k+q}\sigma}^{\dag} \frac{\boldsymbol{\sigma}_{\sigma\sigma^{\prime}}}{2} 
c_{\mathbf{k}\sigma^{\prime}} . \label{eqn2.1.5}
\end{equation}
The Landau quasiparticles are stable low-energy excitations of the Landau Fermi liquid with dynamical properties such as an effective mass and magnetic moment, which are renormalized by the Landau interactions~\cite{Landau1956,Colemanbook}. On the other hand, the collective CDW and SDW fluctuations are not stable excitations since they can easily decay into particles and holes. Again, a particular point to note is that the Landau interactions are assumed to only involve the forward scattering interactions. In fact, a renormalization group study shows that Landau Fermi liquids are stable in 2D and 3D spaces where the Landau forward scattering interactions are indeed marginal~\cite{Shankar1994}.

In summary, Landau Fermi liquids have the following essential properties: (i) the ground state connects adiabatically and analytically to a Fermi gas without any phase transition, and (ii) the low-energy physics is dominated by fermionic quasiparticles which have conserved quantum numbers 
and dynamical properties that are renormalized by the Landau forward interactions.

The above discussion thus shows us how to define Landau Fermi liquids from the perspective of the degrees of freedom. We are then naturally guided toward a new perspective that unconventional electronic physics beyond the Landau Fermi liquid theory framework can be induced by restricting the degrees of freedom of Landau Fermi liquids. These restrictions can be implemented by modifying either the symmetry or the topology with the assistance of external constraints or by interaction-driven dynamical processes.  In particular, we regard the topology as a special global degree of freedom of the quantum states whose nontrivial topological structures may lead to novel physics beyond traditional Landau Fermi liquid theory. The driving mechanisms for the unconventional electronic physics can then be summarized as follows: (i) symmetry breaking, (ii) new emergent symmetries, and (iii) nontrivial topology.  

Indeed, Landau's principle of spontaneous symmetry breaking is a universal mechanism~\cite{Landaubook} that can be used to break down the traditional Landau Fermi liquid theory. Microscopically, the large number of near-zero energy collective fluctuations lead to the instability of the Landau Fermi liquid phase, which then undergoes a spontaneous symmetry breaking phase transition. Accompanying the phase transition, a new macroscopic degree of freedom emerges, the so-called order parameter, whose value is zero in the normal state and which becomes finite in the ordered phase. In comparison to the Landau Fermi liquid phase, the ground state of the ordered phase exhibits a reduced number of symmetries. Examples of such symmetry breaking phase transitions include the SDW, CDW, or superconducting (SC) phase transitions. 
In the SDW phase transition, both the spatial translational and $SU(2)$ spin symmetries are broken, while in the CDW phase transition, the spatial translational symmetry is broken in the particle-hole charge channel.
The SC phase transition, on the other hand, is associated with the spontaneous breaking of the charge $U(1)$ symmetry. In all of these three cases, however, the Landau quasiparticles are gapped in the respective ordered phases. 

As examples of the second driving mechanism, the novel critical phenomena that manifests near the critical point of a continuous phase transition stem from an emergent scaling symmetry, where lots of fluctuations within the divergent correlation length drive the power-law-dependent critical physics~\cite{WilsonRG1974}. In the case of 1D quantum electron systems, a conformal symmetry emerges where infinite conservation laws with dramatic constraints lead to another class of universal states, the so-called Luttinger liquids~\cite{Voit1994}. Similarly, a $U(1)$ gauge symmetry is proposed to emerge in the case of Mott--Hubbard physics, 
where the spin and charge degrees of freedom are strongly correlated, which can lead to many novel physical phenomena beyond the standard Landau Fermi liquid framework~\cite{PALeeRMP2006}. Examples of the nontrivial topology driving mechanism can be found in the quantum Hall effect~\cite{thouless82, avron2003}, and in topological 
insulators~\cite{SchnyderPRB2008,kitaev2009,HasanKane2010,QiZhangRMP2011,Chiu2016}.

\subsection{Analyticity of single-particle Green's functions} \label{sec2.2}

The central concept of Landau Fermi liquid theory is that the low-energy physics is dominated by fermionic quasiparticles that survive the Landau interactions and which have conserved quantum numbers. Mathematically, 
the Landau quasiparticles can be described by the single-particle Green's function
$G_{\sigma} (\mathbf{k},\tau) = - \langle T_{\tau} c_{\mathbf{k}\sigma}(\tau) c_{\mathbf{k}\sigma}^{\dag}(0) \rangle$, 
where $\tau$ denotes imaginary time and $\langle  A \rangle = Tr \left(\rho_H A \right)$ defines the ensemble average, where $\rho_H$ is the density matrix. If the interactions turn on adiabatically without any phase transition, a perturbation formalism can be established and the Dyson equation yields  
\begin{equation}
G_{\sigma}(\mathbf{k},i\omega_n) = \frac{1}{i\omega_n - \varepsilon^{(0)}_{\mathbf{k}\sigma}
-\Sigma_{\sigma}(\mathbf{k},i\omega_n) } , \label{eqn2.2.1}
\end{equation}
where $\Sigma_{\sigma}(\mathbf{k},i\omega_n)$ is the self-energy.

In the case of Landau Fermi liquids, the single-particle Green's function can be separated into coherent and incoherent parts, 
$G_{\sigma}(\mathbf{k},i\omega_n) = G^{(coh)}_{\sigma}(\mathbf{k},i\omega_n) + G^{(inc)}_{\sigma}(\mathbf{k},i\omega_n)$,
where at low energies the coherent part follows  
\begin{equation}
G^{(coh)}_{\sigma}(\mathbf{k},i\omega_n) = \frac{Z_{\mathbf{k}\sigma}}
{i\omega_n - \varepsilon_{\mathbf{k}\sigma} - i \Gamma_{\sigma}(\mathbf{k},i\omega_n)} . 
\label{eqn2.2.2}
\end{equation}
Here, $\varepsilon_{\mathbf{k}\sigma} = \varepsilon^{(0)}_{\mathbf{k}\sigma} + \Re \Sigma_{\sigma}(\mathbf{k},\varepsilon_{\mathbf{k}\sigma})$ and 
$\Gamma_{\sigma}(\mathbf{k},i\omega_n) = Z_{\mathbf{k}\sigma} \Im \Sigma_{\sigma}(\mathbf{k},i\omega_n)$ 
with $Z_{\mathbf{k}\sigma} = \left( 1-\frac{\partial\Re\Sigma}{\partial \omega_n} \right)^{-1} \Big|_{i\omega_n = \varepsilon_{\mathbf{k}\sigma}}$. In order to separate the coherent part of the Green's function from the incoherent part, the self-energy near the Fermi energy is assumed to be analytic. 

One important property of $G^{(coh)}_{\sigma}(\mathbf{k},z)$ is that in the complex $z$-plane, it has a {\em unique pseudo-pole} whose center is located at $\varepsilon_{\mathbf{k}\sigma} + i\Gamma_{\sigma}(\mathbf{k},\varepsilon_{\mathbf{k}\sigma})$. 
In the particular case where the energy is very close to the Fermi energy,
$\Gamma_{\sigma}(\mathbf{k},\varepsilon_\mathbf{k})\rightarrow 0^{+}$, the pseudo-pole becomes a pure pole. $Z_{\mathbf{k}\sigma}$ is the residual weight of the corresponding pole. This single-pole structure describes a well-defined quasiparticle, which is characterized by the energy 
$\varepsilon_{\mathbf{k}\sigma}$, the life-time $1/\Gamma_{\sigma}$, and the coherent weight $Z_{\mathbf{k}\sigma}$. Moreover, the single-pole structure of the single-particle Green's function produces a corresponding peak in the spectrum that can be observed by angle-resolved photoemission spectroscopy.

When this Landau quasiparticle description breaks down, there are some features that may arise in the single-particle Green's function which can be summarized as follows.

\textit{1. Nonanalytic self-energy with $Z_{\mathbf{k}\sigma}\rightarrow 0$}~\cite{VarmaPhysRep2002}: 

Although the Green's function still follows the Dyson equation (\ref{eqn2.2.1}), singular scatterings can also lead to a nonanalytic self-energy, e.g., $\Sigma(\mathbf{k}_{F},\omega) \sim \ln\frac{\omega_c}{\omega}+i|\omega|$, occurring in the case of a marginal Fermi liquid~\cite{VarmaMFL1989,VarmaPhysRep2002}. In such cases, the Landau quasiparticles near Fermi energy lose their coherent weight and this typically leads to the breakdown of the Landau quasiparticles in strong-coupling theories, such as in the study of electrons coupled to spin fluctuations in high-$T_c$ cuprate superconductors~\cite{ChubukovReview2002}. 
     
\textit{2. Multipole singularities}~\cite{Voit1994}:

As an illustrative example, let us consider a SC mean-field state with Hamiltonian 
$H=\sum_{\mathbf{k}} \psi_{\mathbf{k}}^{\dag} \left( 
\varepsilon_{\mathbf{k}}\tau_3 +\Delta_{\mathbf{k}} \tau_1 \right) \psi_{\mathbf{k}}$, 
where $\tau_i$ are the Pauli matrices and 
$\psi_{\mathbf{k}}^{\dag} = \left(c_{\mathbf{k}\uparrow}^{\dag}, c_{-\mathbf{k}\downarrow}\right)$ are Nambu 
spinor operators. The single-particle Green's function is then given by
\begin{equation}
G_{\sigma}(\mathbf{k},i\omega_n)=\frac{\alpha_{\mathbf{k}\sigma}}{i\omega_n - E_{\mathbf{k}}}
+ \frac{\beta_{\mathbf{k}\sigma}}{i\omega_n + E_{\mathbf{k}}} , \label{eqn2.2.3}
\end{equation}
where $E_{\mathbf{k}} = \sqrt{\varepsilon_{\mathbf{k}}^{2} + \Delta_{\mathbf{k}}^{2}}$. Compared to the Green's 
function (\ref{eqn2.2.2}) for Landau Fermi liquids, the Green's function in the SC state has two separate poles, which are symmetrically located on the two sides of the Fermi energy but with different coherent weights
$\alpha_{\mathbf{k}\sigma}, \beta_{\mathbf{k}\sigma}$. 
Physically, this implies that a Bogoliubov quasiparticle in the SC state is the combination of a single particle and hole pair. Similarly, one can show that the single-particle Green's function in the ordered SDW or CDW state also has two separated singular poles. The change from the single-pole singularity structure of Landau Fermi liquids into the multipole structure in the ordered state is another route to the breakdown of the Landau quasiparticle description. We note that the Yang--Rice--Zhang Green's function proposed for high-$T_c$ cuprate superconductors also exhibits separated singular poles~\cite{YangRiceZhang}.

\textit{3. Branch-cut singularities}:

Suppose the Green's function follows
\begin{equation}
G_{\sigma}(\mathbf{k},i\omega_n) = \frac{A}{\left(i\omega_n - 
\varepsilon_{\mathbf{k}\sigma}\right)^{1-a} } , \label{eqn2.4}
\end{equation}  
where the exponent ($a \not= 0$) describes the anomalous dimension of the renormalized electrons. In this case, the analyticity of the Green's function is essentially modified, since the single-pole singularity 
has been transformed into a branch-cut one, indicating the breakdown of the Landau quasiparticle description. This form of branch-cut singularity structure has been identified in several novel electron systems such as 1D Luttinger liquids~\cite{Voit1994}, Fermi liquids in fractal space~\cite{WenXGPRB1993}, Anderson's Hidden Fermi liquid theory~\cite{AndersonHFL2008}, 
and for electrons coupled to scale-invariant unparticles~\cite{Phillips201608}. 

Some additional remarks might be useful to conclude this section. The breakdown of the Landau quasiparticle description is a fundamental characteristic of the breakdown of the Landau Fermi liquid theory framework. Accordingly, substantial modifications of the single-particle Green's function are evident when this breakdown occurs, as discussed above. However, there are instances that may go beyond this scope, e.g., the integer quantum Hall effect. In a more general sense, this may also be the case for topological insulators, where novel topological features typified by edge excitations or surface states emerge as the consequences of nontrivial topology in the quantum states. Despite the lack of symmetry breaking or emergence of new symmetries in such systems, and although the bulk excitations may not be much different from the normal Landau Fermi liquids, these topological states of matter are beyond the reach of a traditional Landau Fermi liquid theory description. Moreover, in the case of interacting topologically nontrivial systems, the interplay between interactions, symmetry, and topology can be much more complicated~\cite{Chiu2016}, and is clearly well beyond the standard Landau theory description. Therefore, we categorize these examples of novel physics beyond the Landau Fermi liquid theory framework as being driven by nontrivial topology.

\section{Restriction of spatial degrees of freedom} \label{sec3}

In this section, we review the effects of restricting the spatial degrees of freedom of quantum electrons. We consider how self-similarity and scaling symmetry at the critical point leads to anomalous dimensions and fractal physics, and review how Fermi liquids in confined 1D space have one stable fixed point, i.e., they form universal Luttinger liquids. We also review how impurity-induced randomness can lead to Anderson localization and the Anderson metal--insulator phase transition.

\subsection{Self-similarity at a critical point: anomalous dimensions and fractal physics} \label{sec3.1}

The basic variable used to describe continuous phase transitions is an order parameter $\phi(\mathbf{r})$, which has the value zero in the normal state and which becomes finite when a spontaneous symmetry breaking induces the continuous phase transition~\cite{Landaubook}. In general, the order parameter is an emergent macroscopic degree of freedom of the continuous phase transition and near the critical point, lots of fluctuations of the order parameter within the divergent correlation length exhibit cooperative critical behaviors. When this occurs, Landau quasiparticles and their microscopic short-range interactions take on secondary roles and the electron system assumes a critical state defined by a single universality class~\cite{WilsonRG1974,FisherRMP1998}. The divergent correlation length and emergent scaling symmetry at the critical point then lead to singular power-law critical behavior of the thermodynamical responses~\cite{MaShangKeng}, and these singular critical behaviors demonstrate self-similarity of the critical phenomena.

The most unusual mystery of the critical phenomena is the finite anomalous dimension $\eta$~\cite{FisherRMP1998,Goldenfeld1992}, which appears in the correlation function of the order parameter,
\begin{equation}
\mathcal{G}(\mathbf{k}) \sim k^{-2 + \eta} , \label{eqn3.1.1}
\end{equation}
where $\mathcal{G}(\mathbf{k}) = \int d^{d}\mathbf{r} \langle \phi(\mathbf{r}) \phi(0) \rangle e^{i \mathbf{k}\cdot \mathbf{r}}$. This finite anomalous dimension stems from the novel correlation of the low- and high-energy fluctuations and violates normal dimensional analysis since the normal dimension of $\mathcal{G}(\mathbf{k})$ should be $-2$. In order to restore the correct normal dimension of $\mathcal{G}$, one can introduce a short-range length $a_l$ so that $\mathcal{G}(\mathbf{k}) \sim  a_l ^{\eta} k^{-2 + \eta}$. This insertion clearly demonstrates the unusual role of short-range fluctuations in critical phenomena~\cite{Goldenfeld1992}. The coupling of electrons to the critical collective fluctuations of the order parameter can lead to the anomalous dimension of the electrons and the breakdown of the Landau quasiparticle description, in similar fashion to the case of electrons coupled to scale-invariant unparticles~\cite{Phillips201608}.

This self-similarity and finite anomalous dimension provides us with a new perspective: the critical phenomena are specific cases of {\em fractal} physics. Near a critical phase transition with an emergent scaling symmetry, electrons are restricted into an effective fractal spatial space which is renormalized by the interactions and has an anomalous dimension. We note that the fractal nature of critical phenomena has also been proposed previously~\cite{Suzuki1983,Kroger2000}, and remark that the anomalous dimension and fractal nature of critical phenomena have a deep relation to an 
underlying nonlinearity. Suppose that a physical variable $A$ with a dimension $d$ of the space ($V\sim L^{d}$) has a scale dimension $d_s= d-\eta$ near a critical point. Under the scale transformation $l^{\prime} = b l$, $A$ becomes $A^{\prime}(L^{\prime}) = b^{-(d-\eta)} A(L)$. Thus, we have $A \sim L^{d-\eta} \sim V ^{\frac{d-\eta}{d}}$, showing that the critical behavior of $A$ is deeply related to nonlinear correlations. This is the 
nonlinear origin of the effective fractal dimension. 

The physical effects of fractal space have been studied by Bares and Wen~\cite{WenXGPRB1993}, for the case where the electrons are confined in the spatial dimension $1<d<2$, and where the Coulomb interaction has fractal power-law behavior. They found non-Fermi liquid physics with a branch-cut singularity in the single-particle Green's function, as per (\ref{eqn2.4}).
While quantum phase transitions and quantum criticality have been demonstrated in many strongly correlated electrons, the nature of the associated spatial and dynamical critical phenomena still remains elusive. We propose that the singular behaviors of the electrons in the quantum critical regime are deeply related to an effective spatial--temporal space with fractal dimension due to the nonlinear interaction effects.   

\subsection{Dimensional confinement: 1D Luttinger liquids} \label{sec3.2}

The most dramatic illustration of the influence of dimensional confinement takes place when the electrons are restricted within a 1D space. On account of the point structure of the resulting Fermi surface, the low-energy 1D electrons have an approximately linear dispersion and this linear dispersion leads to a special spectral structure of the collective SDW and CDW excitations, which are stable and cannot decay into low-energy Landau quasiparticles. Meanwhile, the low-energy Landau quasiparticles are now unstable with respect to the interactions. Thus, the 1D electrons are in the so-called universal Luttinger liquid phase~\cite{Haldane1981}.

In the language of the renormalization group, the Luttinger liquids have a universal low-energy fixed point, the solvable Luttinger model, where only the linear dispersion and the forward interactions are relevant and/or marginal in the renormalization. The nonlinear dispersion and backward and Umklapp scatterings only act to renormalize the Fermi velocities and the effective coupling constants, which are the essential quantities in the definition of the universal properties of the solvable Luttinger model~\cite{Haldane1981,Voit1994}.    

The universal properties of the solvable Luttinger model are as follows~\cite{Voit1994}: (i) the absence of low-energy Landau quasiparticles; (ii) anomalous dimensions of the fermionic operators which produce non-universal power-law correlations; (iii) spin--charge separation; and (iv) universal relations among the non-universal exponents of the correlation functions, and between the renormalized velocities and coupling constants. 
All of these universal properties are deeply rooted to the underlying conformal symmetry which features explicitly in the bosonization of the Luttinger model. The infinite number of local conformal transformations lead to an infinite number of conservation laws, which allows one to determine all the universal properties of the Luttinger model, while the anomalous dimensions of the operators and the exponents of the power-law correlations are associated with the conformal dimensions of the primary fields~\cite{Voit1994}.

\subsection{Impurity-induced Anderson localization} \label{sec3.3}

When impurities are introduced into a quantum electron system, the spatial translational symmetry is broken and electrons may experience emergent Anderson localization ~\cite{AndersonLocalization1958,LagendijkPhysToday}. It was shown in a renormalization group study~\cite{AbrahamsScale1979} that Anderson localization is a universal phenomenon. In fact, in the presence of impurities, all the quantum electronic states in 1D spatial space are localized, while in 2D spatial space, all the quantum electronic states are localized but the localization is marginal. In the case of 3D spatial space, 
things become more complex as some quantum electronic states are localized while others are extended, two of which are separated by the Mott mobility edge $E_c$~\cite{Mott1987}. Thus, depending on the relative values of $E_F$ 
and $E_c$, the electrons can be either in a metallic state $(E_F < E_c)$ or in an insulating state $(E_F > E_c)$. When $E_F = E_c$, the electrons are at a critical point associated with the so-called Anderson metal--insulator 
phase transition. In the Anderson localization mechanism, the breaking of the spatial translational symmetry implies that the momentum is no longer a conserved quantum number, and the quantum electrons are subsequently localized without well-defined Landau quasiparticles.  

Since Anderson's seminal work in 1958~\cite{AndersonLocalization1958}, much effort has been expended on describing the microscopic mechanism of Anderson localization. One scenario is known as {\em weak} localization~\cite{VollhardtLocalization1980,VollhardtLocalization1982} which states that the electron localization arises from quantum interference of the backscattering processes, where one path and its time-reversed partner undergo constructive interference. This quantum interference enhances the backscattering probability and can thus lead to electron localization. As such, in the {\em weak} localization scenario, time-reversal symmetry should be preserved. However, it has been found by Anderson {\em et al.} that the electrons can also be localized in a fractal Cayley-tree lattice where no loops are involved, which indicates that closed loops are not necessary 
for electron localization~\cite{AndersonThouless}. In more modern treatments, Anderson localization remains an interesting and elusive subject, and many new ideas and concepts have been proposed to explain its origin~\cite{EversRMP2008,Roeck2016,LiXPLocalization2016}. 

\section{Restriction of internal degrees of freedom} \label{sec4}

In this section, we investigate the restriction of the internal degrees of freedom of quantum electrons. The following physics will be reviewed and discussed: (i) quantum Hall effects in 2D electronic systems subjected to 
strong magnetic fields with nontrivial topology; (ii) spontaneous symmetry breaking with Fermi-surface instabilities that can lead to phase transitions, e.g., SDW, CDW, Pomeranchuk, or SC instabilities; and (iii) Mott--Hubbard physics with constraints on on-site double occupations.

\subsection{2D electrons subjected to strong magnetic fields: Quantum Hall effects with nontrivial topology} \label{sec4.1}

As has already been mentioned, in this work, we regard the topology of quantum states as a kind of degree of freedom assigned to them. Indeed, from a classification point of view, two systems cannot be deformed into each other in a smooth and continuous way if they are topologically different. In many cases, this kind of distinction is nonperturbative and qualitatively significant. The study of topological issues in condensed matter physics has a long history, from the early investigations of topological defects~\cite{mermin1979} and excitations (see, e.g., Ref.~\onlinecite{kosterlitz1973}), 
to more recent studies of topological quantum states and phase transitions that have attracted intense interest~\cite{HasanKane2010,QiZhangRMP2011,Chiu2016}. 

The quantum Hall effect (QHE) is the prime example of a topologically nontrivial state of matter. When electrons are confined to two dimensions and subjected to a strong magnetic field perpendicular to the plane, 
the Hall conductance exhibits quantized plateaus that are multiples of the elementary conductance, i.e., $\sigma_H=\nu e^2/h$, and this effect can be observed directly at sufficiently low temperatures (For an early review, see e.g., Ref.~\onlinecite{prange2012}). The prefactor $\nu$ is known as the ``filling factor", and is defined as the ratio of the electron density $n$ and the 2D magnetic flux density: $\nu=n\Phi_0/B$, where $B$ is the magnitude of the magnetic field and $\Phi_0=hc/e$ is the flux quantum. The filling factor can take on either integer ($\nu=1,2,3,\ldots$) or particular fractional values such as $\nu=1/3, 2/5, 5/2,\ldots$ etc. The former case is known as the integer quantum Hall effect (IQHE), and was first reported in 1980~\cite{klitzing80}. The latter effect, known as the fractional quantum Hall effect (FQHE), was first observed in 1982 for $\nu=1/3$~\cite{tsui82}, and occurs when the 2D electrons are confined to very high-quality semiconductor interfaces.

The quantization of the Hall conductance in the IQHE can be understood as a consequence of the formation of discrete Landau levels when the electrons are confined to two dimensions with an external magnetic field perpendicular to the plane. For {\em free} electrons in 2D, the quantized Landau levels are equally separated by a ``cyclotron energy" gap ($\hbar\omega_c=\hbar eB/mc$, assuming that the electronic spins are fully polarized), with each level being macroscopically degenerate. The integer-valued filling factor $\nu$ then simply counts the number of filled Landau levels. From this point of view, it is clear that the Landau levels (or the energy bands, if the system is subjected to a lattice-like periodic potential) can be characterized by a topological invariant known as the Chern number. Owing to the 2D nature of the system and the existence of the homogeneous magnetic field which explicitly breaks the time-reversal symmetry, the Chern numbers of the Landau levels are nontrivial~\cite{AvronPRL1983}, and the summation of the Chern numbers of the filled Landau levels turns out to be the Hall conductance in fundamental units~\cite{thouless82, avron2003}. Thus, the appearance of the Hall conductance plateaus is closely related to the discreteness of the topological number, i.e., the quantization of the Hall conductance 
is protected by a topological invariant. The IQHE is the precursor of the later topological classification of free-fermion systems (for a recent review, see e.g., Ref.~\onlinecite{Chiu2016}).

In contrast to the IQHE which can be basically understood on the basis of filled Landau levels with non-interacting electrons, the FQHE (where the Hall conductance plateaus emerge at fractional filling factors) has to be understood by taking electron--electron interactions into account in a more serious manner. In the first place, it is worth pointing out that the FQHE can be regarded as an extreme case of a strongly interacting many-body system. The reason for this is that within a given partially filled Landau level, all the electrons have the same kinetic energy and the Hamiltonian of the system only contains interaction terms and can be written as:
\begin{equation}
	H=\mathcal{P}\sum_{j<k}\frac{e^2}{\ensuremath{\left| {\bf r}_j-{\bf r}_k \right|}}\mathcal{P},
\end{equation}
where $\mathcal{P}$ is the Landau-level projector which projects operators to the limited Hilbert space of the given Landau level (for simplicity we have ignored Landau level mixing effects and the spin degree of freedom). We can see that in this situation, there is no small parameter available and also no ``normal" state to serve as a starting point for traditional perturbative calculations. Moreover, the introduction of the projector $\mathcal{P}$ makes the problem highly nontrivial. 

Fortunately, following the seminal work of Laughlin~\cite{laughlin83}, much progress has been made in understanding the nature of the FQHE. Of particular importance is the concept of the composite fermion, which was first proposed by Jain in 1989~\cite{jain89}, and which provides a clear physical picture for it. A composite fermion is the bound state of an electron and an even number of quantized vortices. These vortices cancel part of the Aharonov--Bohm phases originating from the external magnetic field, which in turn, greatly reduces the strength of the effective magnetic field that the composite fermions experience. Jain further argued that a large part of the original strong interaction between electrons can be reformulated into the internal correlations of the electrons and their binding vortices, with much weaker residual interactions remaining between the composite particles. Thus, to some extent, the composite fermions can be regarded as the ``quasiparticles" of the FQHE. In the framework of composite fermions, we can then map most of the {\em fractional} quantum Hall states of the electrons to the {\em integer} Hall states of the composite fermions, and the various important features of FQHE can be understood quantitatively with the help of numerical simulations~\cite{jainbook}. 

It seems proper to stress here that due to the binding of the electrons to the vortices, these composite particles are collective, nonlocal, and topological quantum particles. The topological nature of the quantum Hall fluids can also be expressed in an effective field theory with a Chern--Simons term included~\cite{zhang89}. Many novel features related to the topological issues, which are well beyond what occur in the generic non-interacting cases, have been identified in these systems (see e.g., Ref.~\onlinecite{Nayak:2008}). The interplay of topology and symmetry 
in many-body systems is currently under intensive examination. 

\subsection{Spontaneous symmetry breaking (I): SDW, CDW, and Pomeranchuk instabilities} \label{sec4.2}

A Fermi surface is called {\em nesting} if a lot of Fermi-surface points are connected by one special momentum and the Fermi surface features parallel regions. The Fermi surface of a Landau Fermi liquid is unstable if it exhibits nesting. With the spin magnetic 
susceptibility defined as $\chi_s(\mathbf{q},\tau) = -\langle T_{\tau}\mathbf{S}_{\mathbf{q}}(\tau) \cdot \mathbf{S}_{-\mathbf{q}}(0) \rangle$ and the SDW operator $\mathbf{S}_{\mathbf{q}}$ defined as in Eq. (\ref{eqn2.1.5}), Fermi surface nesting can lead to the divergence of the static spin susceptibility, i.e., $\chi_s(\mathbf{Q},\omega)\big|_{\omega=0} \rightarrow +\infty$, where $\mathbf{Q}$ is the characteristic 
nesting momentum. This divergence indicates the instability of the Fermi surface in the spin channel. Similarly, a Fermi-surface instability can also occur in the charge channel. With the CDW operator $n_{\mathbf{q}}$ defined as in Eq. (\ref{eqn2.1.4}), we can define the charge density susceptibility as $\chi_c(\mathbf{q},\tau) = -\langle T_{\tau} n_{\mathbf{q}}(\tau)\cdot n_{-\mathbf{q}}(0)\rangle $, and the nesting Fermi surface can also lead to the divergence of the static charge susceptibility, $\chi_c(\mathbf{Q},\omega)\big|_{\omega=0} \rightarrow +\infty$. Which one of the SDW or CDW instabilities is preferred is determined by the degrees of the divergence of the two susceptibilities.

The Fermi-surface instabilities outlined above can induce a spontaneous symmetry breaking phase transition below a critical temperature, where along with this phase transition, quantum electrons go from a normal Landau 
Fermi liquid state into an ordered state. In the case of a SDW phase transition, the spatial translational symmetry and the spin $SU(2)$ symmetry are both broken, and the SDW order parameter $\langle \mathbf{S}^{z}_{\mathbf{Q}} \rangle$ becomes finite. In the CDW phase transition, the spatial translational symmetry in the particle--hole charge 
channel is broken, and a finite CDW order parameter $\langle n_{\mathbf{Q}} \rangle$ emerges. In the ordered SDW or CDW state, the ground state can be regarded as the condensation of particle--hole pairs in the spin or charge channels with SDW or CDW Goldstone excitations. Moreover, the Landau quasiparticles are gapped and their single-particle Green's function exhibits a pair of singular poles, as highlighted in Sec. \ref{sec2.2}. Microscopically, the nesting Fermi surface becomes unstable when there are finite interactions whose transferred momentum is the Fermi-surface nesting momentum $\mathbf{Q}$. These types of interactions with finite momentum $\mathbf{Q}$ are beyond the Landau forward scattering interaction description.  

There is another Fermi-surface instability known as the Pomeranchuk instability~\cite{Pomeranchuk1959}. Suppose the Landau interaction in the charge channel has the form
\begin{equation}
H_I = - U_{\rho} \sum_{\mathbf{k}\mathbf{k}^{\prime}} \phi_{\mathbf{k}} \phi_{\mathbf{k}^{\prime}} 
\bar{n}_{\mathbf{k}} \bar{n}_{\mathbf{k}^{\prime}} , \label{eqn4.2.1}
\end{equation}
where $\bar{n}_{\mathbf{k}} =\sum_{\sigma} c_{\mathbf{k}\sigma}^{\dag} c_{\mathbf{k}\sigma}$. Then, when $U_{\rho}$ is positive and larger than a critical value $U_{\rho}^{(c)}$, there will be a Pomeranchuk deformation of the Fermi surface with the deformation symmetry determined by $\phi_{\mathbf{k}}$. Similarly, if the Landau interaction in the spin channel follows
\begin{equation}
H_I = - U_s \sum_{\mathbf{k}\mathbf{k}^{\prime}} \phi_{\mathbf{k}} \phi_{\mathbf{k}^{\prime}} 
\bar{\mathbf{S}}_{\mathbf{k}} \cdot  \bar{\mathbf{S}}_{\mathbf{k}^{\prime}} , \label{eqn4.2.2}
\end{equation}
where $\bar{\mathbf{S}}_{\mathbf{k}} =\sum_{\sigma\sigma^{\prime}} c_{\mathbf{k}\sigma}^{\dag} 
\frac{\boldsymbol{\sigma}_{\sigma\sigma^{\prime}}}{2} c_{\mathbf{k}\sigma^{\prime}}$, and $U_s$ is positive 
and larger than a critical value $U_s^{(c)}$, a spin-dependent Pomeranchuk Fermi surface deformation may occur. In contrast to the SDW and CDW Fermi surface instabilities that occur for finite momentum $\mathbf{Q}$, the Pomeranchuk Fermi surface deformation can be regarded as a zero-momentum instability. 

The Pomeranchuk Fermi surface deformation is a particular manifestation of electronic nematicity and breaks the rotational symmetry of the system~\cite{Fradkin}. Thus, the Pomeranchuk--Landau interactions (\ref{eqn4.2.1}) and (\ref{eqn4.2.2}) can drive a phase transition that spontaneously breaks the rotational symmetry and thus result in an electronic nematic state with a finite nematic order parameter $\sum_{\mathbf{k}}\langle \phi_{\mathbf{k}} \bar{n}_{\mathbf{k}}  \rangle$ or 
$\sum_{\mathbf{k}}\langle \phi_{\mathbf{k}} \bar{\mathbf{S}}_{\mathbf{k}}  \rangle$. 
At the first glance, the electronic nematic state appears to be normal, similar to anisotropic Landau Fermi liquids; however, this is only a superficial mean-field manifestation. The scattering interactions between the Landau quasiparticles and the Goldstone nematic modes of the rotational symmetry breaking mechanism can lead to unusual non-Fermi liquid behavior 
of the self-energy of the Landau quasiparticles, and simultaneously, of the damped Goldstone nematic  modes~\cite{Fradkin,FradkinPRB2001,WatanabeVishwanath}. These novel scattering physical effects are also preserved in the case of a nematic phase transition in a discrete crystal with rotational symmetry breaking, where it is found that the discrete lattice effects are irrelevant over most of the temperature range~\cite{FradkinPRB2001}. Thus, the electronic nematic 
state is another example of a non-Fermi liquid state. Further novel physics associated with nematic criticality and non-Fermi liquid physics also occurs near the critical point of a nematic phase transition~\cite{Fradkin,FradkinPRB2001,LawlerFradkin2006}.

\subsection{Spontaneous symmetry breaking (II): Superconductivity} \label{sec4.3}

Landau Fermi liquids have a global charge $U(1)$ symmetry under phase transformations of the form $c_{i\sigma}\rightarrow c_{i\sigma}e^{i\theta}$, which implies the conservation of the particle number. When the global charge $U(1)$ symmetry is broken, Landau Fermi liquids transform into a SC state following a phase transition in which the superconductivity emerges.

In the SC state, the SC order parameter 
\begin{equation}
\widehat{\Delta} = \sum_{\mathbf{k}\sigma\sigma^{\prime}}\Delta_{\sigma\sigma^{\prime}}(\mathbf{k})
c^{\dag}_{\mathbf{k}\sigma} c^{\dag}_{-\mathbf{k}\sigma^{\prime}} \label{eqn4.3.1}
\end{equation}
has a finite average value, and neglecting spin-orbit coupling, we can decouple the spin part into its spin singlet and spin triplet components. In the spin singlet SC state,
$\Delta_{\sigma\sigma^{\prime}}(\mathbf{k}) = \left[-i\Lambda(\mathbf{k})\sigma_2\right]_{\sigma\sigma^{\prime}}$
with $\Lambda(-\mathbf{k}) = \Lambda(\mathbf{k})$, while in the spin triplet state,
$\Delta_{\sigma\sigma^{\prime}}(\mathbf{k}) = \left[-i\left(\mathbf{d}(\mathbf{k})\cdot \boldsymbol{\sigma}\right)
\sigma_2\right]_{\sigma\sigma^{\prime}}$
with $\mathbf{d}(-\mathbf{k}) = -\mathbf{d}(\mathbf{k})$~\cite{SigristUedaRMP1991}.
The finite pairing order parameter reflects the macroscopic condensation of Cooper pairs, where this macroscopic condensation stems from the fixed phase of the Cooper pairs which contribute coherently to the whole condensate. Collective phase modes and amplitude modes can then be found over the Cooper pair condensate ground state, and on account of the coupling of the charged superconductor to electromagnetic gauge fields, the 
longitudinal phase modes become high-energy plasmon modes~\cite{AndersonPlasmon1958}. One well-known example of such transverse 
and topological phase modes are vortices, while the amplitude modes of the SC condensate, known as SC Higgs modes~\cite{VarmaHiggsPRL1981,VarmaHiggsPRB1982}, 
are still being sought for in experiments. As mentioned in Sec. \ref{sec2.2}, the Landau quasiparticles are gapped in the SC state and their single-particle Green's functions exhibit a double-pole singularity.

We note that the SC state can also be regarded as an example of the low-energy Fermi-surface nesting instability in the particle-particle channels 
with nesting momentum $\mathbf{Q}=0$ (the momentum of the center of mass of Cooper pairs). It has been shown by Shankar's renormalization group study that the SC instability is relevant to Landau Fermi liquids, and that the SC 
state is a stable fixed point of the electron liquid~\cite{Shankar1994}.

\subsection{Gutzwiller projection: Mott--Hubbard physics} \label{sec4.4}

When the $d$- or $f$-electron shells of ions are not fully filled, strong residual on-site Coulomb interactions known as Mott--Hubbard interactions can heavily suppress on-site double occupations. The relevant physics of such systems can be described by the Hubbard model and involves both kinetic electron hopping terms and their on-site Mott--Hubbard interactions. The strong suppression of on-site double occupations can be described by the 
Gutzwiller projection~\cite{Gutzwiller1963}
\begin{equation}
\mathcal{P} = \prod
_{i} \left( 1- \alpha n_{i\uparrow} n_{i\downarrow} \right) . \label{eqn4.4.1}
\end{equation}  
Generally, one takes $0\leq \alpha \leq 1$ with $\alpha=0$ ($\alpha=1$) expressing the absence of the reduction (exact exclusion) 
of on-site double occupations. The Gutzwiller projection is thus a restriction on the local basis states of the Hilbert space, where the local on-site doubly-occupied basis states are reduced or even completely removed. 

In particular, the Gutzwiller projection can lead to a Mott--Hubbard metal--insulator phase transition when the $d$- or $f$-electron 
shells are half-filled. When virtual hopping processes are included, one then finds that there are effective low-energy super-exchange antiferromagnetic interactions, and as a result, a Mott--Hubbard insulator can show long-range antiferromagnetic order in quasi-2D or 3D compounds at low temperatures. When additional electrons or holes are introduced into such half-filled compounds, the long-range antiferromagnetic order is gradually broken and the system evolves into a spin-liquid or spin-glass state. Moreover, superconductivity can occur when the motion of the doped electrons or holes becomes coherent. 

In 1987, Anderson proposed the resonating valence bond (RVB) scenario to describe the superconductivity of hole-doped high-$T_c$ cuprate superconductors~\cite{Anderson1987}, where the Gutzwiller projection plays a key role in the description of their low-energy physics. In this account, the superconductivity emerges when the holes exhibit coherent motion in the RVB spin-liquid background, and thus form a novel non-Bardeen--Cooper--Schrieffer (non-BCS) mechanism for the superconductivity. Anderson's RVB 
proposal has been extensively studied by many analytical and numerical methods, where in order to deal with the local Gutzwiller 
projection, a $U(1)$ gauge theory~\cite{PALeeRMP2006} is established with the electron operator $c^{\dag}_{i\sigma}$ reformulated as
\begin{equation}
c^{\dag}_{i\sigma} = f^{\dag}_{i\sigma} b_{i} , \label{eqn4.4.2}
\end{equation}
and where the spin and the charge degrees of freedom of the electrons are separately carried by spinons $f_{i\sigma}$ and holons $b_i$, respectively. This decoupling is based upon an assumption that the Landau quasiparticles are not well-defined in Mott--Hubbard physics, but rather related to spin--charge separation. There is then an emergent $U(1)$ gauge symmetry 
with one additional gauge degree of freedom that leads to strong correlations between the spinons and the holons. Novel physics such as non-BCS superconductivity, the pseudogap, and the strange metallic state (among other examples) can then arise. We note that there are various different gauge theories proposed for the Gutzwiller projection constraint with similar assumptions and similar outputs. However, the brilliant RVB proposal and the various proposed gauge theories are still in doubt, as the predicted spinons and holons have not been found in experiments.  

Mott--Hubbard physics has also been extensively studied by wavefunction Gutzwiller projection methods~\cite{VollhardtRMP1984,EdeggerGros2007}. In such descriptions, the Gutzwiller projected state is defined as 
$\big| \Psi \rangle_{P} = \mathcal{P} \big| \Phi_0 \rangle$, where $\big| \Phi_0 \rangle$ is a non-interacting Fermi gas or a mean-field BCS SC state. The Gutzwiller projected Fermi gas turns out to be a renormalized Fermi 
liquid~\cite{VollhardtRMP1984}, while the Gutzwiller projected BCS state is a renormalized BCS state~\cite{LiTao2007,Yunoki2006,EdeggerGros2007},
both of which have quasiparticle excitations with finite coherent weight. These results are in stark contrast to the original Anderson's RVB scenario where the electrons should be in non-Fermi liquid states and exhibit spin--charge separation. Whether such coherent Landau quasiparticles can survive the strong Mott--Hubbard interactions and the question of how the unconventional superconductivity emerges from the pseudogap state or the strange metallic state are still big puzzles in the modern field of condensed matter.

The key role, however, of Mott--Hubbard interactions is their reduction of on-site double occupations. This reduction is thus a restriction of the local basis states of the Hilbert space and obviously breaks the Pauli exclusion principle of the Landau quasiparticles. This restriction may then lead to strong correlations between the spin and the charge degrees of freedom, such as those driven by strong $U(1)$ gauge fluctuations, as 
proposed in $U(1)$ gauge theory~\cite{PALeeRMP2006}. Since the Pauli exclusion principle is a fundamental principle associated with the Fermi--Dirac statistics of Landau quasiparticles, the breaking of the local Pauli exclusion principle may break the Fermi--Dirac statistics of the Landau quasiparticles and consequently lead to novel excitations with emergent new statistics.

\section{Restriction of external degrees of freedom} \label{sec5}

As alluded to earlier in this manuscript, Landau Fermi liquids only involve forward scattering Landau interactions, while 
renormalization group studies have shown that the Landau forward scattering interactions are marginal and Landau Fermi liquids are thus stable~\cite{Shankar1994}. In this section, we will discuss electron systems that are coupled to external degrees of freedom. The models we will consider for review include electrons coupled to Goldstone bosons, Anderson's hidden Fermi liquid, electrons coupled to scale-invariant unparticles, and Kondo physics and heavy fermion superconductors where the electrons are coupled to local magnetic moments. 

\subsection{Electrons coupled to Goldstone bosons: Watanabe--Vishwanath theorem} \label{sec5.1}

Recently, Watanabe and Vishwanath outlined a theorem for the stability of Landau Fermi liquid behavior and Goldstone bosons in metals exhibiting spontaneous symmetry breaking~\cite{WatanabeVishwanath}. In the case of a 
continuous phase transition with spontaneous symmetry breaking, we may define generators $\{Q_{a}\}$ which are associated with the Goldstone modes. We then consider the commutation relations 
\begin{equation}
\left[ Q_a, P_i \right] = i \Lambda_{a i} , \label{eqn5.1.1}
\end{equation}   
where $P_i$ is the $i$-th component of the momentum $\mathbf{P}$. If $\Lambda_{a i} = 0$, the coupling between the electrons and the Goldstone modes vanishes in the limit of the small energy--momentum transfer, while if 
$ \Lambda_{a i} \not= 0$, the electron--Goldstone-boson coupling does not vanish. Since vanishing coupling preserves the stability of the Landau Fermi liquid and the Goldstone modes, a non-vanishing coupling may lead to novel non-Fermi liquid behavior and the breakdown of the Goldstone modes. The Watanabe--Vishwanath theorem thus shows scattering restriction effects on Landau Fermi liquids from the collective modes in the channel of the spontaneous symmetry breaking.  

According to the theorem, since the coupling between the electrons and the phonons in crystal compounds vanishes at small momentum transfer, both the Landau Fermi liquid states and the phonons are stable. This is similar to when electron--magnon coupling vanishes in ferromagnetic metals, in which case the magnons are stable. Generally speaking, all Goldstone modes associated with internal symmetry breaking are stable; however, nonvanishing coupling between electrons and Goldstone nematic modes arising from spontaneous rotational symmetry breaking may lead to non-Fermi liquid physics and overdamped Goldstone nematic modes. Non-Fermi liquids exhibiting rotational symmetry breaking are thus subject to extensive study~\cite{WatanabeVishwanath,Fradkin,FradkinPRB2001,LedererKivelson2015}. 
We note also that the coupling between electrons and transverse gauge bosons does not vanish, which can lead to novel non-Fermi liquid behavior at extremely low temperatures~\cite{TsvelikBook2003,WatanabeVishwanath}.
    
\subsection{Anderson's hidden Fermi liquid} \label{sec5.2}

Recently, Anderson proposed a hidden Fermi-liquid theory for the high-$T_c$ cuprate superconductors~\cite{AndersonHFL2008,AndersonHFL2009}. For the projected $t$--$J$ model with Hamiltonian $H_{tJ}$, there is a one-to-one correspondence between the exact eigenstates $\big| \Psi \rangle$ and the unprojected states $\big| \Phi \rangle$, i.e., 
\begin{equation}
\big| \Psi \rangle = \mathcal{P} \big| \Phi \rangle , \label{eqn5.2.1}
\end{equation}  
where $\mathcal{P}$ is the Gutzwiller projection with exact exclusion of the double occupations defined in 
(\ref{eqn4.4.1}). It can be shown that if $H_{tJ}\big| \Psi \rangle = E \big| \Psi \rangle$, then 
$H_{tJ}\big| \Phi \rangle = E \big| \Phi \rangle$. $\big| \Phi \rangle$ describes the non-interacting Fermi-gas states or the mean-field BCS states. Since $\hat{c}^{\dag}_{i\sigma}\equiv \mathcal{P} c^{\dag}_{i\sigma}\mathcal{P} = c^{\dag}_{i\sigma} (1-n_{i\bar{\sigma}})$, 
the single-particle Green's function of the ground state 
$G_{ij}(t) = - i \langle \Psi_G \big| T_t c_{i\sigma}(t) c^{\dag}_{j\sigma}(0) \big|\Psi_G \rangle$
can be simplified to
\begin{equation}
G_{ij}(t) \sim G^{(0)}_{ij}(t) G^{(\star)}_{ij}(t) , \label{eqn5.2.2}
\end{equation} 
where $G^{(0)}_{ij}(t) = - i \langle \Phi_G \big| c_{i\sigma}(t) c^{\dag}_{j\sigma}(0) \big| \Phi_G \rangle $ 
describes the free electron propagation, and 
$G^{(\star)}_{ij}(t) = \langle \Phi_G \big| T_t (1-n_{i\bar{\sigma}})(t) (1-n_{j\bar{\sigma}})(0) \big|\Phi_G \rangle $ 
describes the background fluctuations scatterings off the quasiparticles. $\big|\Psi_G \rangle$ and $\big|\Phi_G \rangle$ 
are the correlated ground states defined by (\ref{eqn5.2.1}).

With this simplification, Anderson showed that~\cite{AndersonHFL2008} 
\begin{equation}
G(\mathbf{k},\omega) \sim \frac{1}{\left( \omega - v_F k \right)^{1-a}} , \label{eqn5.2.3}
\end{equation}
where $v_F$ is the Fermi velocity. The finite exponent $a$ describes the anomalous dimension of the electron operators arising from the singular background scatterings, and has a dramatic effect since it leads to a branch-cut singularity in the single-particle Green's function, an obvious characteristic of the breakdown of the Landau quasiparticles. Physically, the breakdown of the Landau quasiparticles originates from the novel scattering effects of the background fluctuations, which are deeply related to the on-site double occupation constraint. In similar fashion to our previous discussion in Sec. \ref{sec4.4}, this may be deeply rooted in an underlying emergent gauge symmetry, as proposed in the RVB scenario~\cite{PALeeRMP2006} and in the breakdown of the Fermi--Dirac statistics. We note that the hidden Fermi liquid scenario for high-$T_c$ cuprate superconductors can also be extended to describe 2D electrons with quantum Hall effects~\cite{JainAnderson}.

\subsection{Electrons coupled to unparticles} \label{sec5.3}

Limtragool {\em et al.} have introduced a phenomenological model to describe the power-law singular behavior of the strange metallic state of high-$T_c$ cuprate superconductors~\cite{Phillips201608}. They attribute the power-law behavior to scale-invariant critical physics whose generation involves the coupling of electrons to bosonic unparticles which have scale-invariant features. These unparticles have novel propagators 
\begin{equation}
\mathcal{D}_{\mu} \left(\mathbf{q}, \nu_n \right) = 
\frac{1}{\left( \nu_n^{2} + E_{\mathbf{q}}^{2} \right)^{1-\alpha} } , \label{eqn5.3.1}
\end{equation}
where the exponent $\alpha= 1-\frac{d+1}{2}+d_{\mu}$, $d$ is the spatial dimension, and $d_{\mu}$ is the scaling dimension of the unparticle field.  Since the Green's function of the unparticles exhibits a branch-cut 
singularity rather than a general pole singularity, these unparticles are not generally well-defined bosons. It can be shown that at low temperatures and energies, the scattering from the scale-invariant unparticles leads to 
non-Fermi liquid behavior of the electron self-energy, 
\begin{equation}
\Im \Sigma \sim T^{d-2+2\alpha}, \big| \omega \big|^{d-2+2\alpha} . \label{eqn5.3.2}
\end{equation}

The role of these scale-invariant unparticles is to introduce unusual scattering restrictions on the electrons in the particle--hole charge channel. Physically, these bosonic unparticles with scale-invariant features may arise 
from the high-energy degrees of freedom of the electrons, and contribute to the scaling-invariant critical-like physics. The branch-cut singularity in the unparticle propagator then plays a similar role to the power-law Coulomb 
interaction in fractal space~\cite{WenXGPRB1993} in the formation of the non-Fermi liquid behaviors.

\subsection{Conduction electrons coupled to local moments: Kondo effects and heavy fermions} \label{sec5.4}

When conduction electrons are coupled to local magnetic impurities or local $f$-electrons, novel physics can emerge, such as in the case of universal Kondo effects and heavy fermions with unconventional superconductivity. 

When magnetic impurities are introduced into a metallic conductor, certain universal Kondo effects beyond the Landau Fermi-liquid theory can arise~\cite{ColemanbookCh17}, including: (i) a resistivity minimum, (ii) universality of the specific heat $C_V\sim \frac{1}{T} f_c \left(\frac{T}{T_K}\right)$, and (iii) universality of the resistivity $\rho(T)\sim f_{\rho}\left(\frac{T}{T_K}\right)$. 
The Kondo temperature $T_K$ is the unique characteristic energy scale for the universal Kondo effects, and the universal Kondo effects outlined above can be described by the Kondo-impurity model, 
\begin{equation}
H = \sum_{\mathbf{k}\sigma} \varepsilon_{\mathbf{k}} c^{\dag}_{\mathbf{k}\sigma}
c_{\mathbf{k}\sigma} + J \sum_{\sigma\sigma^{\prime}} c^{\dag}_{i_0\sigma} 
\frac{\boldsymbol{\sigma}_{\sigma\sigma^{\prime}}}{2} c_{i_0\sigma^{\prime}} 
\cdot \mathbf{S}_{f} , \label{eqn5.4.1}
\end{equation}
where $\{c, c^{\dag}\}$ are the operators of the conduction electrons and $\mathbf{S}_{f}$ is the impurity spin 
located at site $i_0$. Such Kondo effects result from the universal physics of the Kondo resonance stemming from the local coupling of the conduction electrons to the magnetic impurity. It should be noted that the Kondo 
coupling $J$ has a dramatic character known as asymptotically free behavior, i.e., its renormalized value is small at high energy but is large at low energy~\cite{AndersonPoorman}. This shows that the Kondo resonance is a particular instance of criticality.

When the local magnetic moments form a crystalline lattice configuration, such as for $f$-electrons in heavy fermion superconductors, a more suitable model for their description is the Kondo-lattice model,
\begin{equation}
H = \sum_{\mathbf{k}\sigma} \varepsilon_{\mathbf{k}} c^{\dag}_{\mathbf{k}\sigma}
c_{\mathbf{k}\sigma} + J \sum_{i\sigma\sigma^{\prime}} c^{\dag}_{i\sigma} 
\frac{\boldsymbol{\sigma}_{\sigma\sigma^{\prime}}}{2} c_{i\sigma^{\prime}} 
\cdot \mathbf{S}_{f,i} . \label{eqn5.4.2}
\end{equation}  
One special feature of the unconventional physics in heavy fermion superconductors is the emergence of heavy fermions with a heavily renormalized mass ($m^{\star}/m \sim 200$). Physically, these heavy fermions are analogues of the Kondo resonance and form a nearly flat band near the Fermi energy. 

Both the Kondo resonance and heavy fermions can be regarded as composite fermions arising from the binding of the conduction electrons and the spin-flip scattering of local magnetic moments. Mathematically, the composite fermions can be defined by~\cite{ColemanbookCh17}
\begin{equation}
\sum_{\sigma^{\prime}} \left( \boldsymbol{\sigma} \cdot 
\mathbf{S}_{f,i}(t)\right)_{\sigma\sigma^{\prime}} c_{i\sigma^{\prime}}(t^{\prime}) 
\rightarrow \Delta(t-t^{\prime}) f_{i\sigma}(t^{\prime}), \label{eqn5.4.3}
\end{equation}
where $f_{i\sigma}$ is the operator of the composite fermions, and $\Delta(t-t^{\prime})$ is the analogue of the Cooper pair wavefunction. These composite fermions are spatially localized but are strongly correlated in dynamic 
processes~\cite{SiQMLocalQCP2003}, and highly nonlocal in the temporal space, thus forming {\em local} Fermi liquids~\cite{NozieresLocalFL1973}. In this scenario, the Kondo resonance and the heavy fermions can be regarded 
as emergent composite fermions resulting from the Kondo-coupling driven time correlations between the conduction electrons and the local magnetic moments.  

While the Kondo effects of the magnetic impurity induced Kondo resonance are well understood, the complex correlations of heavy fermions are far beyond our knowledge. One central problem in the description of heavy fermion 
superconductors is the dynamical creation and destruction of heavy fermions. Physically, this problem is equivalent to the localization and delocalization of the $f$-electrons which are hybridized strongly with the conduction electrons. These central problems involve the competition of the Kondo liquid of the heavy fermions and the spin liquid of the local moments, where the former involves local Kondo couplings and the latter involves Ruderman--Kittel--Kasuya--Yosida (RKKY) interactions between the local moments. A simple phenomenological two-fluid model has been proposed to account for the basic physics of heavy fermion superconductors~\cite{YangYFPines2012,YangYFPines2014}; 
however, the driving mechanisms for the complex competing physics are still little known. Examples of such competing physics include phase transitions between the Fermi-liquid and non-Fermi-liquid states of heavy fermions, the heavy fermion Kondo liquid, the spin liquid, and the ordered magnetic state of the local moments. A particular challenge in contemporary condensed matter physics is to understand such quantum phase transitions and associated quantum criticality, as exhibited in most heavy fermion superconductors~\cite{StewartNFLRMP2001,ColemanReview2015,LohneysenRMP2007}.

\section{Summary and proposals} \label{sec6}

\subsection{Summary} \label{sec6.1}

\begin{table*}[thb] 
\caption{ Novel electronic physics arising from the restriction of the degrees of freedom (DOF) of Landau Fermi liquids. 
Notation and abbreviations are defined as follows: B/I space -- spatial--temporal/internal space, 
$\mathbf{r}$- or $t$-space, $\mathscr{H}$ -- Hilbert space, sym. -- symmetry, 
emgt. -- emergent, brk. -- breaking, stats. -- statistics,  ST -- spatial translational, 
FD -- Fermi--Dirac. } \label{tab6.1}
\begin{ruledtabular}
\begin{tabular}{llll}
novel physics & DOF & B/I space  &  symmetry or topology \\
\hline
critical phenomena  & scale  & B ($\mathbf{r}$)  & emgt. scaling sym.  \\
Luttinger liquids & dimension & B ($\mathbf{r}$)  & emgt. conformal sym. \\
Anderson localization & translation    & B ($\mathbf{r}$) & ST sym. brk.  \\
QHE & topology  & I ($\mathscr{H}$)  & emgt. nontrivial topology \\
SDW & translation \& spin   & B ($\mathbf{r}$) \& I (spin) & ST and spin $SU(2)$ sym. brk.  \\
CDW & translation \& charge  & B ($\mathbf{r}$) \& I (charge) & ST sym. brk. in charge channel  \\
Pomeranchuk & rotation \& spin/charge  & B ($\mathbf{r}$) \& I (spin/charge) & rotational sym. brk. in spin/charge channel \\
SC & $U(1)$ phase & I (phase) &  charge $U(1)$ sym. brk. \\
Mott--Hubbard physics & double occupations & I ($\mathscr{H}$)  & emgt. gauge sym., FD stats. brk. \\
\begin{tabular}{l}
electrons with  \\
Goldstone modes
\end{tabular}
 & Goldstone modes   & B ($\mathbf{r}$) \& I   &  sym. brk. \\
hidden Fermi liquid &  double occupations  & I ($\mathscr{H}$) &  emgt. gauge sym. (underlying) \\
\begin{tabular}{l}
electrons with  \\
unparticles
\end{tabular}
 & unparticles  & I (charge) & scale invariance of unparticles \\
\begin{tabular}{l}
electrons with  \\
local moments
\end{tabular} 
&  local moments  & B ($\mathbf{r},t$) \& I (spin)  & emgt. FD stats. of composite fermions 
\end{tabular}
\end{ruledtabular}
\end{table*}

In the last several decades, many strongly correlated electron systems have been discovered in experiments, which clearly reveal unconventional electronic physics beyond the traditional Landau Fermi-liquid theory. A major challenge in the modern field of condensed matter is to establish suitable theories to understand these unconventional electronic phenomena. In this article, we have presented a new perspective regarding such unconventional electronic physics that provides us with basic means of studying the breakdown of the Landau Fermi liquid theory, and which serves as a guideline for achieving manual control of electrons for novel physics. 

In this article, we firstly reviewed the essential properties of Landau Fermi liquids: Landau Fermi liquids connect adiabatically and analytically to a non-interacting Fermi gas, their ground state preserves all the symmetries of the system, and their low-energy physics is dominated by Landau fermionic quasiparticles with conserved quantum numbers and whose dynamical properties are renormalized by Landau forward scattering  interactions~\cite{Landau1956,Colemanbook}.
We then presented our perspective that most examples of unconventional electronic physics beyond the Landau Fermi-liquid theory paradigm stem from the restriction of the degrees of freedom of the quantum electrons in Landau Fermi liquids. Since the degrees of freedom of the quantum electrons correspond to independent ``motions" that can be used to identify the quantum states, they can be defined by a complete set of commuting observables. As a complete set of commuting observables can be constructed in terms of the symmetry group of the quantum electrons, the degrees of freedom are thus deeply connected to the symmetry. In addition to the degrees of freedom defined by symmetry, another degree of freedom that may be required to identify the quantum states is the global topology, which can be out of reach of the symmetry-based description and which can have significant physical consequences. With these results, we were naturally led to propose that unconventional electronic physics can be realized by the restriction of the degrees of freedom of the quantum electrons in Landau Fermi liquids via the 
following mechanisms: (i) symmetry breaking, (ii) the emergence of new symmetry, and (iii) nontrivial topology.   

We then extensively reviewed various unconventional electronic physics beyond the Landau Fermi liquid theory framework from our perspective. The considered examples of novel electronic physics arising from the restriction of the degrees of freedom of electrons in Landau Fermi liquids are summarized in Table (\ref{tab6.1}). Novel physics arising from the restriction of spatial degrees of freedom include anomalous dimensions and critical phenomena, Luttinger liquids, and Anderson localization. The anomalous dimensions of critical phenomena near critical points stem from an emergent scaling symmetry~\cite{WilsonRG1974,FisherRMP1998}, where lots of fluctuations within the divergent correlation length lead to singular power-law critical behavior. The values of the critical exponents are deeply connected to the scale-invariant transformations and the anomalous dimensions exhibited in these systems implies the fractal nature of the critical phenomena, and moreover, that the electrons are restricted to an effectively fractal space. In the second example we considered, confinement of the quantum electrons in 1D spatial space results in the formation of 1D Luttinger liquids~\cite{Haldane1981}, whose universal low-energy physics stems from an underlying conformal symmetry of the solvable Luttinger model, i.e., a stable fixed point of Luttinger liquids. The underlying conformal symmetry (with the associated infinitely many conservation laws) plays a crucial role in determining the universal properties of the 1D Luttinger liquids~\cite{Voit1994}. The third example we reviewed was Anderson localization, which arises from impurity-induced randomness which breaks the spatial translational symmetry~\cite{AndersonLocalization1958,LagendijkPhysToday}. We note that in the weak localization scenario, Anderson localization is argued to result from the quantum interference of the backscattering processes which are preserved by time-reversal symmetry~\cite{VollhardtLocalization1980,VollhardtLocalization1982}. 

Next, we considered examples where the internal degrees of freedom are restricted, including the integer and fractional quantum Hall effects; spontaneous symmetry breaking with SDW, CDW, Pomeranchuk, and SC phase transitions; and Mott--Hubbard physics. The quantum Hall effects occur when confined 2D electrons are subjected to a strong magnetic field, and the quantization of the Hall conductance stems from the nontrivial topological structure of the quantum states~\cite{thouless82,avron2003}. In particular, highly topologically nontrivial features emerge in the FQHE case when electron--electron interactions are properly taken into account for partially filled Landau levels. We then noted that when the Fermi surface exhibits nesting, finite interactions at the nesting momentum can drive SDW or CDW phase transitions in the particle--hole spin or charge channels. In the case of a SDW phase transition, both the spatial translational symmetry and the spin $SU(2)$ symmetry are broken, while in the CDW phase transition case, the spatial translational symmetry is broken in the charge channel. Moreover, we highlighted that the ground state of the SDW or CDW ordered state has less symmetry than the normal Fermi-liquid state, and that the SDW or CDW Goldstone excitations are gapless while the Landau quasiparticles are gapped. The next example we considered was when the rotational symmetry of Landau Fermi liquids is broken, in which case a Pomeranchuk nematic phase transition will occur in the spin or charge channels, accompanied by a Pomeranchuk Fermi-surface  deformation~\cite{Pomeranchuk1959,Fradkin}. Interactions between the Landau quasiparticles and the Goldstone nematic modes have unusual influences which can lead to non-Fermi liquid behavior and overdamped Goldstone nematic modes. In the SC case, if the global charge $U(1)$ symmetry of the Landau Fermi liquids is broken, there is macroscopic condensation of Cooper pairs and superconductivity emerges. We highlight the point that in each of the above examples of phase transitions with spontaneous symmetry breaking, new macroscopic degrees of freedom (the so-called order parameters) emerge. We then turned our attention to Mott--Hubbard physics, which is deeply related to the Gutzwiller constraint of strong suppression of on-site double occupations. This is a constraint on the local Hilbert space which can lead to novel correlations between the spin and charge degrees of freedom. In order to explain such non-Fermi liquid behavior, an emergent $U(1)$ gauge symmetry was proposed in Anderson's RVB scenario, providing an additional $U(1)$ gauge degree of freedom~\cite{PALeeRMP2006}. The strong 
suppression of on-site double occupations breaks the Pauli principle and thus leads to the breakdown of the Landau quasiparticles with Fermi--Dirac statistics description. 

Finally, we reviewed examples where unconventional electronic physics can arise from the restriction of external degrees of freedom, such as in the case of Goldstone modes and continuous symmetry breaking, the background fluctuations of Anderson's hidden Fermi liquid, unparticles with scale-invariant features, and local moments in Kondo physics and heavy fermion superconductors. We saw that if the generator of the broken continuous symmetry does not commute with the momentum operator, interactions between the Landau quasiparticles and the Goldstone modes will lead to the breakdown of the Landau Fermi liquid and overdamped Goldstone modes~\cite{WatanabeVishwanath}. 
In the Anderson's hidden Fermi liquid case~\cite{AndersonHFL2008,AndersonHFL2009}, the Gutzwiller constraint on 
on-site double occupations is simplified by scattering effects of the background fluctuations, which leads to the singular propagation of the electrons with anomalous dimension. In the next example, unparticles with scale-invariant features were introduced to study the strange metallic state of the high-$T_c$ cuprate superconductors~\cite{Phillips201608}. Their coupling to electrons was noted to lead to power-law non-Fermi liquid behavior. In our final review, universal Kondo effects were seen to stem from the Kondo resonance arising from the local coupling of the conduction 
electrons and magnetic impurities; however, the complex competing physics accompanying quantum phase transitions and quantum criticality in heavy fermion superconductors remains beyond our knowledge. Such effects have been 
proposed~\cite{YangYFPines2012,YangYFPines2014} to be deeply related to the dynamical creation and destruction of so-called heavy fermions ---being composite fermions formed by the coupling of conduction electrons and the local moments---along with the competition between the heavy fermion Kondo liquid phase and the spin liquid of the local moments.    

\subsection{Proposals} \label{sec6.2}

In our study of the various examples of novel non-Fermi liquid physics, we have focused on the degrees of freedom of particles and taken the physical states as our starting point. Since a degree of freedom of a particle defines a special independent ``motion" of its physical states, in principle, fundamental interactions dependent on this degree of freedom may be fixed.  For example, since the spin degree of freedom stems from the reconciliation of quantum mechanics and the special theory of relativity, the most fundamental spin-relevant interactions are defined as~\cite{DiracBook} 
\begin{equation}
\boldsymbol{\sigma} \cdot \left( \mathbf{p} + e \mathbf{A} \right) . \label{eqn6.1}
\end{equation} 
Thus, we have two fundamental couplings for spin, $\boldsymbol{\sigma} \cdot \mathbf{p}$
which is the origin of the so-called spin-orbit couplings, and $\boldsymbol{\sigma} \cdot \mathbf{A}$
which defines the coupling of the spin to electromagnetic gauge fields. These fundamental interactions guide us in how to detect and manually control the given degree of freedom of the particles. Moreover, focusing on the degrees of freedom of the particles can reveal basic relevant scattering processes, and consequently, some basic physical properties. However, it should be noted that there may be other interactions beyond the fundamental ones. For example, in the case of spin, the direct- and super-exchange spin--spin interactions come from Coulomb interactions and virtual hopping processes due to large on-site Hubbard interactions, respectively.

Nevertheless, since the degrees of freedom of quantum electrons are deeply connected to the overall symmetry and topology, we may take our guiding principles for the description of novel electronic physics to be as follows:
(i) to specify what the symmetry and topology imply for a given degree of freedom and to study the relevant symmetry breaking, emergent symmetry, or nontrivial topology in detail;  
(ii) to specify what the interactions imply for a given degree of freedom and to study the relevant interaction-driven physical properties; 
(iii) to specify the coupling between different degrees of freedom and to study the relevant physical competition or cooperation.

In the case of the breakdown of the Landau Fermi liquid picture, since Landau Fermi liquids are protected by three principal 
hypotheses---namely, (i) the Fermi--Dirac statistics, (ii) the adiabatic and analytic principle, and (iii) the 
quasiparticle-dominated low-energy physics---we can focus on further additional guidelines to account for the breakdown of their description:
(i) How can the Fermi--Dirac statistics of the Landau quasiparticles be broken? Which degrees of freedom are the most relevant to account for this breakdown? 
(ii) What are the possible phase transitions associated with each degree of freedom? 
(iii) How can the Landau quasiparticles be broken down, e.g., causing the well-defined propagator of Landau quasiparticles to break down by the vanishing of the residual weight $Z_{\mathbf{k}\sigma}$, by the singular self-energy, or by an anomalous dimension and branch-cut singularity? 
(since the Landau interactions only involve forward scatterings, we can focus on interactions in other channels, backward scattering, Umklapp scattering, or other singular scatterings induced by collective fluctuations or external degrees of freedom), and 
(iv) To study the possible symmetry breaking, new emergent symmetries, and/or nontrivial topology of the degrees of freedom of the Landau Fermi liquids involved.   

This perspective on the restriction of the degrees of freedom can also be taken as a guideline for the innovative design of electronic devices with novel physical properties.  In particular, we could manually control the electrons in artificial devices by manipulating any of the degrees of freedom, such as the spin, charge, phase, spatial and temporal degrees of freedom, or the topology. Electrons in a confined spatial space with nontrivial topology or with manually controlled artificial potentials may have unusual, interesting, and useful properties.  
A particular point to note, however, is that time may prove to be a special parameter whose role has not been particularly well studied. Dynamical correlations and other possible interaction-driving temporal restrictions may perhaps be relevant to some non-Fermi liquid physics, and may indeed be the case for a proper description of quantum criticality. One of the most dramatic difficulties in the study of many unconventional electronic physical non-Fermi liquid systems is rooted in determining many-body correlations. The question of the basic variables and study of the emergent 
degrees of freedom for many-body correlations are relevant problems for future studies.

{\it Acknowledgements:}
We thank Tao Li and Yin Zhong for helpful discussions. H.T. Lu acknowledges the support of the Natural Science Foundation of China (Grant No. 11474136). Y.H. Su was partially supported by the Natural Science Foundation of 
China (Grant No. 11774299) and the Natural Science Foundation of Shandong Province (Grant No. ZR2017MA033).


%


%

\end{document}